  \documentclass[useAMS,usegraphicx, usenatbib]{mn2e}
\usepackage{amsmath}
\usepackage{amssymb}
\usepackage[utf8x]{inputenc}

\newcommand{\msun}{\,\text{M}_\odot}
\newcommand{\nG}{\,\text{nG}}
\newcommand{\muG}{\,\mu\text{G}}

\begin{document}

\title{Random Primordial Magnetic Fields  and the  Gas Content of  Dark Matter Haloes}
\author[R. S. de Souza, L. F. S. Rodrigues and R. Opher]
{Rafael S. de Souza$^{1}$\thanks{Email: rafael@astro.iag.usp.br}, Luiz Felippe S. Rodrigues$^{1}$\thanks{Email: felippe@astro.iag.usp.br},  and Reuven Opher$^{1}$\thanks{Email: opher@astro.iag.usp.br}
\\ $^{1}$IAG, Universidade de S\~{a}o Paulo, Rua do Mat\~{a}o 1226, Cidade
 Universit\'{a}ria, CEP 05508-900, S\~{a}o Paulo, SP, Brazil\\
}

\date{Accepted -- Received  --}

\pagerange{\pageref{firstpage}--\pageref{lastpage}} \pubyear{2010}

\maketitle
\label{firstpage}

\begin{abstract}
We recently predicted the existence of  random primordial magnetic fields (RPMF) in the form of randomly oriented cells with dipole-like structure with a cell size $L_0$ and an average magnetic field $B_0$. 
Here we investigate models for  primordial magnetic field with a similar web-like structure, and other geometries,  differing perhaps in $L_0$ and $B_0$.  
The effect of RPMF on the formation of the first galaxies is investigated. The filtering mass, $M_F$, is the halo mass below which baryon accretion is severely depressed. 
We show that these RPMF could influence the formation of galaxies by altering the filtering mass and the baryon gas fraction of a halo, $f_g$. 
The effect is particularly strong in small galaxies.
We find, for example, for a comoving $B_0=0.1\muG$, and a reionization epoch that starts at $z_s=11$ and ends at $z_e=8$, for $L_0=100\,\text{pc}$ at $z=12$, the $f_g$ becomes severely depressed for $M<10^7\msun$, whereas for $B_0=0$ the $f_g$ becomes severely depressed only for much smaller masses, $M<10^5\msun$. 
We suggest that the observation of $M_F$ and $f_g$ at high redshifts can give information on the intensity and structure of primordial magnetic fields.
\end{abstract}


\section{Introduction}

Understanding the details of galaxy formation remains an important challenge in cosmology. 
As shown by numerical calculations, the first generation of galaxies should have formed at very high redshifts inside collapsing halos, starting at $z \sim 65$, corresponding to high peaks of the primordial dark matter (DM) density field  \citep{NNB}. Cosmic Microwave Background (CMB) radiation observations  suggest that reionization began at high redshifts.
This means that a high abundance of luminous objects must have existed at that time, since these first luminous objects are expected to have heated and reionized their surroundings
\citep{rev,WL03,HH03,Cen}.

The formation of a luminous object inside a halo necessarily  requires the existence of baryonic gas there. Even in halos that are too small for cooling via atomic hydrogen, the gas content can have substantial, and observable, astrophysical effects. In addition to the possibility of hosting astrophysical sources, such as stars, small halos may produce a 21-cm signature \citep{Kuhlen,Shapiro06,NB08, Furlanetto06}, and can block ionizing radiation and produce an overall delay in the global progress of reionization \citep{bl02, iliev2, iss05, mcquinn07}. 
  
The evolution of the halo gas fraction at various epochs of the Universe is of prime importance, particularly in the early Universe. We  evaluate here the possible influence of a primordial magnetic field on the halo gas fraction. 

As noted by  \citet{Gnedin2000a, Gnedin1997}, both in the linear and non-linear regimes, the accretion of gas into DM halos is suppressed below a characteristic mass scale called the filtering mass, $M_F$. This mass scale coincides with the Jeans mass, $M_J$, if the latter does not vary in time. Otherwise, $M_F$ is a time average of $M_J$. Thus, an increase in the ambient pressure  in the past, causes an increase in $M_J$ and suppresses the accretion of baryons into DM halos in a cumulative fashion, producing an increase in $M_F$.

Until now, studies focused on the UV heating of the neutral interstellar gas as the main source of pressure, for determining  the  filtering mass. These results  are widely used in many semi-analytic models (e.g.  \citealt{Maccio2009}), particularly those designed to study the properties of small galaxies (due to the high redshift character of the UV heating).

In this paper we add the effect of a possible random primordial magnetic field as another important source of ambient pressure. The magnetic field contributes to pressure support, which changes the Jeans mass and, consequently, the filtering mass and the quantity of gas that is accreted by DM halos. 

The paper is organized as follows. In section \ref{sec:magnetic} we make a short review on the possible origins of primordial magnetic fields, in section \ref{sec:filtering} we analyze the effect of primordial magnetic fields on the Jeans and filtering  masses and in section \ref{sec:gas} we calculate effects on the baryon mass fraction. In section \ref{sec:conc} we give our conclusions. 

\section{Primordial Magnetic Fields} \label{sec:magnetic}

The origin of large-scale cosmic magnetic fields in galaxies and protogalaxies remains   a challenging problem  in astrophysics \citep{Zweibel1997, Kulsrud2008, Souza2008, souza10,Souza2010, Widrow2002,lagana}. Understanding the origin of the structures of the present Universe requires a knowledge of the origin of magnetic fields. The magnetic fields fill interstellar and intracluster space and affect the evolution of galaxies and galaxy clusters. 
There have been many attempts to explain  the origin of cosmic magnetic  fields.
One of the most popular astrophysical theories for creating seed primordial fields is that they were generated by the Biermann mechanism \citep{Biermann1950}. It has been suggested that this mechanism acts in diverse astrophysical systems, such as  large scale structure formation \citep{Peebles1967, Rees1972, Wasserman1978},  protogalaxies \citep{Davies00}, cosmological ionizing fronts \citep{Gnedin2000a},  star formation and  supernova explosions  \citep{Hanayama2005, Miranda1998}. Another mechanism for creating cosmic magnetic fields was suggested by \citet{Ichiki2006}. They  investigated the second-order couplings between photons and electrons  as a possible origin of  magnetic fields on cosmological scales before the epoch of recombination.  Studies of magnetic field generation,   based
on cosmological perturbations, have also been made  \citep{Takahashi2005, Takahashi2006, Clarke2001, Maeda2009}.

In our galaxy, the magnetic field is coherent over kpc scales with alternating directions 
in the arm and inter-arm regions (e.g., \citealt{Kronberg1994, han08}). Such alternations are expected for magnetic fields of primordial origin \citep{Grasso2001}.
 
Various observations put upper limits on  the intensity of  a homogeneous primordial magnetic field. Observations of the small-scale cosmic microwave background (CMB) anisotropy yield an upper comoving limit of $2.98\nG$ for a homogeneous primordial field \citep{Yamazaki2010}. Reionization of the Universe puts upper limits of $\sim 0.7-3 \nG$ for a homogeneous primordial field, depending on the assumptions of the stellar population that is responsible for reionizing the Universe \citep{Schleicher2008}. 

\citet{Souza2008,Souza2010} suggested that the fluctuations of the plasma predicted by the Fluctuation Dissipation Theorem,  after the quark-hadron transition (QHT), is a natural source for a present primordial magnetic field. They evolved the fluctuations after the QHT to the present era and predict a present cosmic web of random primordial magnetic fields. The average magnetic field predicted by them over a region of size $L$ is $B = 9 \muG ~(0.1 \text{ pc}/L)^{3/2}$.  An average magnetic field $0.003\nG$ over a $2$~kpc region at $\emph{z} \sim 10$ is, thus, predicted.

\section{Effects on the filtering mass}\label{sec:filtering}

\subsection{The filtering scale}
Following the procedure of a previous work \citep{Rodrigues2010}, which studied the effects of a homogeneous primordial magnetic field, we study here the influence of random inhomogeneous primordial magnetic fields (RPMF) on the filtering mass $M_F$. This quantity describes the highest DM mass scale for which the baryon accretion is suppressed significantly, as we will discuss below.

First, we define the filtering scale \citep{Gnedin1997} as the characteristic length scale over which the baryonic perturbations are smoothed out as compared to the dark matter ones  as
\begin{equation}
 \frac{\delta_b}{\delta_{tot}}=1-\frac{k^2}{k_F^2} \,\text{ ,}
\end{equation}
where $\delta_b$ is the density contrast of baryonic matter and $\delta_{tot}$, the total density contrast. For $k$ comparable to $k_F$, the density contrast $\delta_b$ is severely depressed.

As was shown by \citet{Gnedin2000a}, we can relate the comoving wavenumber associated with this length scale with the Jeans wavenumber by the equation 
\begin{equation}
 \frac{1}{k^2_F(a)}=\frac{3}{a}\int^a_0 \frac{da'}{k_J^2(a')} \left[ 1-\left(\frac{a'}{a}\right)^\frac{1}{2} \right],\label{eq:kj}
\end{equation}
where a flat matter dominated universe is assumed. 

One finds that the overall suppression of the growth of baryonic density perturbations depends on a time-average of the Jeans scale. By translating the length scales into mass scales, we can then define the Jeans mass and filtering mass,
\begin{equation}
  M_J \equiv \frac{4\pi}{3} \bar{\rho} \left(\frac{2\pi a}{k_J}\right)^3\,\,\text{ and }\,\, M_F \equiv \frac{4\pi}{3} \bar{\rho} \left(\frac{2\pi a}{k_F}\right)^3 \,\text{ .}
  \label{eq:MJMF}
\end{equation}

From equations (\ref{eq:MJMF}) and (\ref{eq:kj}), we can write,
\begin{equation}
 M_F^\frac{2}{3}=\frac{3}{a} \int^a_0 da'\,\, M_J^\frac{2}{3}(a')\left[ 1-\left(\frac{a'}{a}\right)^\frac{1}{2}\right]\text{ ,}\label{eq:filtering}
\end{equation}
where $\bar \rho$ is the mean matter density.

The commonly used Jeans mass, with negligible magnetic fields, is the mass when the gravitational pressure at the surface of a sphere of radius $R_J$ balances the thermal pressure. An adiabatic compression of the sphere by a change in radius $\delta R$ increases the thermal pressure above the gravitational pressure, causing the sphere to increase its radius and oscillate  about the equilibrium value $R_J$.

When the thermal pressure is negligible and we only have random magnetic fields in the sphere, the definition of the Jeans mass is similar.  It is the mass when the magnetic pressure at the surface balances the gravitational pressure. An adiabatic compression of the sphere of radius $R_J$ by a change in radius $\delta R$ increases the magnetic pressure above the gravitational pressure, making the sphere, again, increase its radius and oscillate about the radius $R_J$. 

\subsection{Magnetic fields and pressure}

For a random magnetic field,  the magnetic pressure in a region of comoving size $L$ greater than the comoving size of a magnetic cell, $L_0$,     is given by  \citep{Hindmarsh1998}
\begin{equation}P=\frac{B_{rms}^2}{8\pi}, \label{Prms}
\end{equation}
with the following expression for the rms average of the field \citep{Grasso2001,Souza2008} 
\begin{equation}
B_{rms}(a)=\sqrt{\langle B^2\rangle}=B_{0}\left(\frac{L_{0}}{L}\right)^{p}\left(\frac{a_{0}}{a}\right)^{2},\label{eq:Brms}
\end{equation}
where $B_0$  is the field intensity in an individual cell, and the parameter $p$ depends on the geometry of the field considered (section \ref{sec:cell}).

For $L<L_0$, the average is being made inside a single cell. Thus, the field is indistinguishable from a homogeneous field \citep{Rodrigues2010}, and we have
\begin{equation}
 B_{rms}(a) = B_{0}\left(\frac{a_{0}}{a}\right)^{2}\text{ .} \label{eq:Brms2}
\end{equation}

\subsection{Turbulence}
\label{sec:turb}

Equations (\ref{eq:Brms}) and (\ref{eq:Brms2}) can be improved taking into account the turbulent enhancement of $B$ at large length scales, which occurs until $B$ reaches equipartition with the kinetic energy of the plasma. An inverse cascade effect occurs, where small magnetic structures merge to form larger magnetic structures, transferring energy to larger length scales.
Numerical simulations  suggest that the total enhancement can be written as $f_{T}(t) \simeq e^{t/\tau}$, where $\tau$ is the eddy turn over time of the intergalactic turbulence. The mean value of $\tau$ is   $\tau \sim 10^9$ years \citep{Ryu2008}. 

Thus, equations (\ref{eq:Brms}) and (\ref{eq:Brms2}) become
\begin{align}
\langle B^2\rangle &=  f^2_{T}(z)B^2_0\left(\frac{L_0}{L}\right)^{2p}\left(1+z\right)^{4}& \text{ for } L>L_0\text{ ,} \label{eq:B1}\\
\langle B^2\rangle &= f^2_{T}(z)B_0^2 \left(1+z\right)^4 & \text{ for }  L<L_0\text{ .}\label{eq:B2}
\end{align}

When the field reaches equipartition, the turbulent amplification stops. To take into account  this effect in our calculations, we set an upper limit to the magnetic field of $B\approx 0.1\muG $ for the comoving strength of the field when averaged over 1 kpc. This is consistent with the expected values for magnetic fields in equipartition with the environment in regions around clusters and groups \citep{Ryu2008}.

This also consistent with tests that we made stopping the amplification when $\langle B^2\rangle\sim 8\pi\,\rho k T$.
\subsection{Obtaining the Jeans mass}

It is to be noted that it is not the Alfvenic speed, determined by $B_0$, which sets the timescale for an overdensity to respond to perturbations. A simple example shows this. Let a perturbation be made along the magnetic  field, $B_0$, in a given cell on the surface of the sphere. In that cell the Alfvenic speed is determined by $B_0$. Let us assume that the perturbation enters a neighboring cell that could have its field $B_0$  perpendicular to the direction of propagation of the perturbation. In this neighboring cell the Alfven velocity of the perturbation is zero since Alfvenic perturbations can propagate only along the field.
From the above example, we conclude that in a sphere of randomly oriented cells, the velocity of perturbations is not the Alfven velocity defined by $B_0$, but is determined by the average magnetic pressure determined by $B_{rms}^2 (\ll B_0^2)$.

We are interested in obtaining the appropriate Jeans wave number, $k_J$, for a sphere of radius $L$ containing randomly oriented magnetic cells of size $L_0$ with average magnetic fields $B_0$. The usually used $k_J$, when magnetic fields are negligible, is $k_J=a\sqrt{4\pi G\rho}/c_s$, where $c_s$ is the speed of sound. In such a sphere, the speed of sound sets the timescale for an overdensity to respond to perturbations, and is directly related to the pressure. 

In a sphere with a homogeneous magnetic field, $B_H$, the speed of a perturbation propagating perpendicular to $B_H$ is $v_{ma}=\sqrt{\frac{B^2_H}{4\pi\rho} + c_s^2}$, the magneto-acoustic velocity, which sets the timescale. The energy density in the sphere is $\frac{B^2_H}{4\pi}$.

In our case of random magnetic fields, the average energy density in the sphere is $\frac{B^2_{rms}}{4\pi}$. We may, then, expect that the characteristic velocity in our sphere of random magnetic cells is approximately given by the expression for the magneto-acoustic velocity given above, with the energy density $\frac{B^2_H}{4\pi}$ replaced by $\frac{B^2_{rms}}{4\pi}$. Defining an effective Alfven velocity by $\bar v_A^2 = \frac{B^2_{rms}}{4\pi\rho}$, the characteristic velocity of a perturbation in our sphere is, then, $v_c=\sqrt{c_s^2+\bar v_A^2}$.

Replacing $c_s$ by $v_c$ (in the usual expression for $k_J$ when there is negligible magnetic fields) we then have
\begin{equation}
 \frac{k_J}{a}=\left(\frac{4\pi G\rho}{c_s^2+\bar v_A^2}\right)^{1/2},
\end{equation}
which we use in this paper.

Thus, the Jeans mass of a plasma, subject to magnetic pressure, is given by
\begin{equation}
M_{J}^2=\frac{3}{4\pi G^{3}\bar \rho}\left(\frac{B_{rms}^{2}}{4\pi \bar \rho}+\frac{3}{2}\frac{k_{B}T}{m_{H}\mu}\right)^{3}  \,\text{ ,}\label{eq:Jeans}
\end{equation}
where we use $c_s=\sqrt{\gamma k_B T / (\mu m_H) }$,  with  $m_H$  being the mass of a hydrogen atom, $\mu$ the mean molecular weight and $k_B$  the Boltzmann constant.

This expression generalizes previous calculations of the Jeans mass which only considered its limiting cases: $B\rightarrow 0$, the usual Jeans mass (e.g. \citealt{Padmanabhan3}),  or $T\rightarrow 0$, the magnetic Jeans mass (e.g.  \citealt{Tashiro2005}). 

In order to choose the correct $B_{rms}$ from either equation (\ref{eq:Brms}) or (\ref{eq:Brms2}), we first calculate the (comoving) Jeans length, $L_m$, from equations (\ref{eq:Brms2}) and (\ref{eq:Jeans}), in which we assume a multi-cell regime

\begin{equation}
L_m^6=\left(\frac{\kappa}{G}\right)^3  \left[\frac{\kappa \,f^2_{T}(z)B_0^2}{3} \left(\frac{L_0}{L_m}\right)^{2 p}  +\frac{3}{2}\frac{k_B T(z)}{\mu m_H}(1+z)^{-1} \right]^3\text{ .}\label{eq:Lm}
\end{equation}
where \(\kappa\equiv\frac{2\, G}{\Omega_{m0} H_0^2}\) and we used
\begin{equation*}
 \bar \rho= \Omega_{m0} \frac{3 H_0^2}{8\pi G} (1+z)^3 =\frac{3}{4\pi} \frac{(1+z)^3}{\kappa}, 
\end{equation*}
 and 
\begin{equation*}
 L_m^3 =\frac{M_J}{\frac{4}{3}\pi \bar \rho}(1+z)^3=\kappa\, M_J\text{ .}
\end{equation*}

If $L_m>L_0$, then $L_m$ is the 
comoving Jeans length and the Jeans mass is given by the solution of
\begin{equation}
 M_J^2=\frac{\kappa}{G^3}
\left[ \frac{\kappa^{\left(1-\frac{2}{3} p\right)}}{3} \frac{f^2_{T}(z) \,B_0^2 L_0^{2 p}}{M_J^{\frac{2}{3} p}}
+\frac{3}{2}\frac{k_B T(z)}{\mu m_H(1+z)} \right]^3\text{.}
\label{eq:Jeansfinal}
\end{equation}

If $L_m<L_0$, then the average is done inside a single cell, using equation (\ref{eq:Brms2}), and the Jeans mass is given by
\begin{equation}
 M_J^2=\frac{\kappa}{G^3}\left[ \frac{\kappa}{3} \,f^2_{T}(z)B_0^2  +\frac{3}{2}\frac{k_B T(z)}{\mu m_H(1+z)} \right]^3
\label{eq:Jeanshom}\text{ .}
\end{equation}

\subsection{Random magnetic field models}
\label{sec:cell}

We study primordial magnetic fields in the form of randomly oriented cells considering two possible scenarios for the seed field.

\paragraph*{Dipole like fields} 

The first scenario we discuss is one where each cell contains a dipole field whose flux is conserved. In this case we have $p=3/2$  \citep{Hindmarsh1998,Souza2008,Souza2010} in equations (\ref{eq:Lm}) and (\ref{eq:Jeansfinal}).

\paragraph*{Ring-like fields}

We also consider the geometry studied by \citet{Ahonen1997} and \citet{Enqvist1993}, who found cells with large ring-like fields, but with planes of inclination randomly 
oriented. Thus, 
an average over large volumes  corresponds to a random walk of all possible inclinations. This is equivalent a random walk on a 2D surface of a sphere, which implies $p=1$.

\subsection{Temperature}

In order to calculate the Jeans and filtering masses from equations (\ref{eq:filtering}) and (\ref{eq:Jeansfinal}), it is necessary to have an expression for the evolution of the temperature of the gas with redshift. We use the analytic fit of the temperature as a function of redshift that  \citet{Kravtsov2004} obtained for the results of \citet{Gnedin2000a},
\begin{equation}
T(z)=\left\{ \begin{array}{ccl}
(10^4\text{ K})\left(\frac{1+z_s}{1+z}\right)^\alpha &\text{,}& z>z_s\\
10^4\text{ K} &\text{,}& z_s \geq  z \geq z_r\\
(10^4\text{ K})\left(\frac{1+z}{1+z_r}\right)  &\text{,}& z<z_r\\
\end{array}\right.\label{eq:temp}
\end{equation}
where $z > z_s$ is the epoch before the first HII regions form, $z_r \leq z \leq z_s$ is the epoch of the overlap of multiple HII regions and  $z < z_r$ is the epoch of complete reionization.

Throughout this paper we use $\alpha=6$, $z_s=11$ and $z_r=8$, unless otherwise mentioned.

\subsection{Results}

\begin{figure}
\includegraphics[width=0.9\columnwidth]{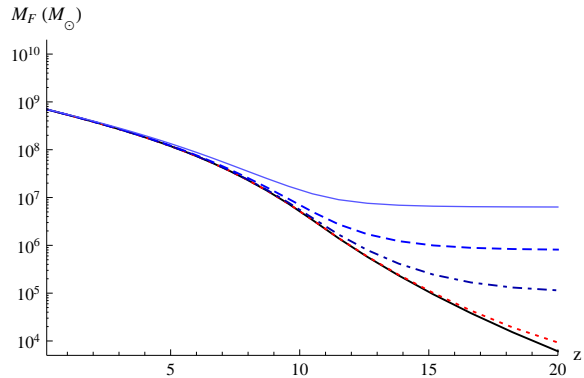}
\caption{Variation of the filtering mass with redshift in the presence of a dipole-like ($p=3/2$) random magnetic field, for $z_s=11$ and $z_r=8$. The continuous (\textit{black}) curve corresponds to the $B_0=0$ case. The other curves have $B_0=0.1\,\mu\text{G}$ and, from bottom to top, 
$L_0=10\text{ pc}$ for the dotted (\textit{red}) curve;
$L_0=10^2\text{ pc}$ for the dash-dotted (\textit{dark-blue}) curve;
$L_0=10^{2.5}\text{ pc}$ for the dashed (\textit{blue}) curve;
$L_0=10^{3}\text{ pc}$ for the thin (\textit{light-blue}) curve.}
\label{fig:filter7}
\end{figure}

\begin{figure}
\includegraphics[width=0.9\columnwidth]{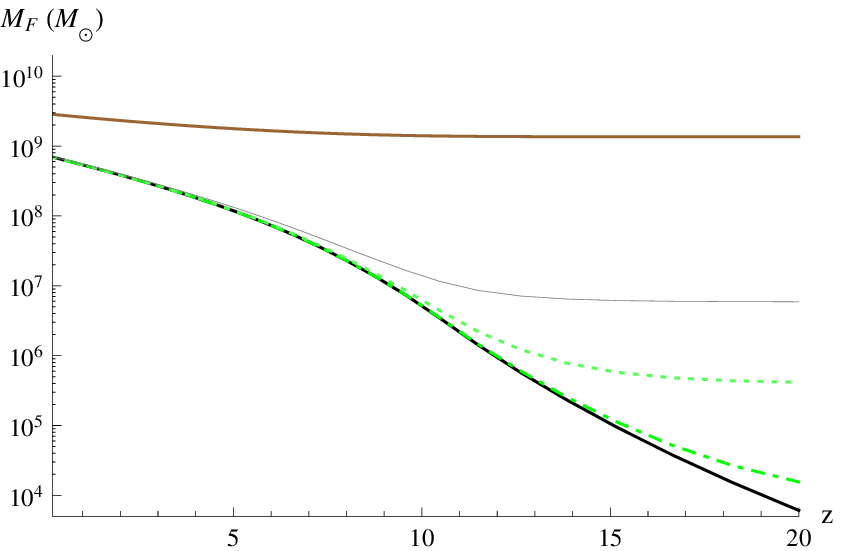}
\caption{
Variation of the filtering mass with redshift in the presence of a  dipole-like  ($p=3/2$) random magnetic field, for $z_s=11$ and $z_r=8$. The bottom continuous (\textit{black}) curve corresponds to the $B_0=0$ case. The top continuous (\textit{brown}) curve corresponds to the CMB upper-limit $B_0\approx2.98\nG$ and $L_0=1\text{ Mpc}$. The other curves have $B_0=10\nG$ and, from bottom to top, 
$L_0=10^{2}\text{ pc}$ for the dash-dotted  (\textit{green}) curve;
$L_0=10^{3}\text{ pc}$ for the dotted (\textit{light-green}) curve;
$L_0=10^{4}\text{ pc}$ for the thin (\textit{gray}) curve.
}
\label{fig:filter8}
\end{figure}

\begin{figure}
\includegraphics[width=0.9\columnwidth]{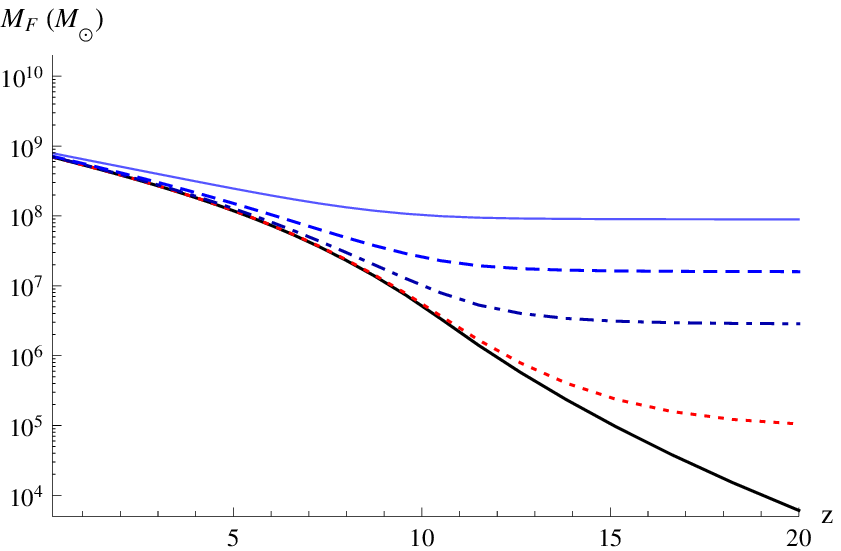}
\caption{ 
Variation of the filtering mass with redshift in the presence of a ring-like ($p=1$) random magnetic field, for $z_s=11$ and $z_r=8$. The continuous (\textit{black}) curve corresponds to the $B_0=0$ case. The other curves have $B_0=0.1\,\mu\text{G}$ and, from bottom to top, 
$L_0=10\text{ pc}$ for the dotted (\textit{red}) curve;
$L_0=10^2\text{ pc}$ for the dash-dotted (\textit{dark-blue}) curve;
$L_0=10^{2.5}\text{ pc}$ for the dashed (\textit{blue}) curve;
$L_0=10^{3}\text{ pc}$ for the thin (\textit{light-blue}) curve.}
\label{fig:filter7ring}
\end{figure}

\begin{figure}
\includegraphics[width=0.9\columnwidth]{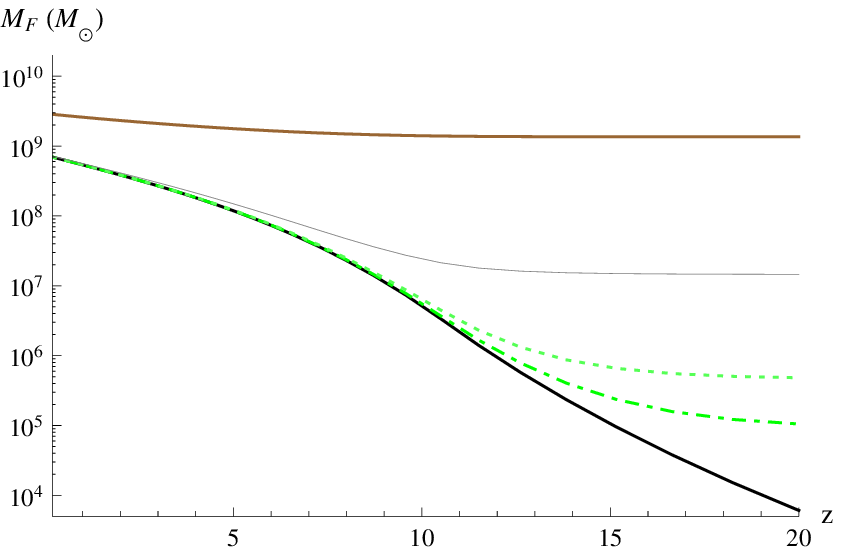}
\caption{
Variation of the filtering mass with redshift in the presence of a ring-like ($p=1$) random magnetic field, for $z_s=11$ and $z_r=8$. The bottom continuous (\textit{black}) curve corresponds to the $B_0=0$ case. The top continuous (\textit{brown}) curve corresponds to the CMB upper-limit $B_0\approx2.98\nG$ and $L_0=1\text{ Mpc}$. The other curves have $B_0=10\nG$ and, from bottom to top, 
$L_0=10^{2}\text{ pc}$ for the dash-dotted  (\textit{green}) curve;
$L_0=10^{3}\text{ pc}$ for the dotted (\textit{light-green}) curve;
$L_0=10^{4}\text{ pc}$ for the thin (\textit{gray}) curve.
}
\label{fig:filter8ring}
\end{figure}

We use equations (\ref{eq:Jeansfinal}), (\ref{eq:Jeanshom}) and (\ref{eq:temp}) in (\ref{eq:filtering}) to calculate the effect of RPMF on the filtering mass. The results obtained by assuming different values for $L_0$ and $B_0$ are shown in figures \ref{fig:filter7} and \ref{fig:filter8}, for dipole-like fields, and in figures \ref{fig:filter7ring} and \ref{fig:filter8ring} for ring-like fields (without taking into account the effects of amplification, i.e. setting  $f^2_{T}(z)\approx 1$).

The model proposed by \citet{Souza2008} leads to dipole-like field with a comoving $B_0\approx 0.1\muG$ and $L_0\approx 1\text{ pc}$. 
This curve deviates only slightly from the case of no magnetic field, 
in figure \ref{fig:filter7}.
We found that most models where magnetic fields are generated during a quark-hadron phase transition -- which would have dipole-like fields with $B_0\approx 2\times 10^{-17}\text{ G}$ and $L_0\approx 1\text{ A.U.}$ \citep{Hogan1983}, or $B_0\approx 10^{-16}\text{ G}$ and $L_0\approx 1\text{ pc}$ \citep{ChengOlinto} -- or during an electroweak phase transition -- ring-like fields with $B_0\sim 10^{-7}$ to $10^{-9}\text{ G}$ and $L_0\sim10\text{ A.U.}$ \citep{Baym1996} -- have negligible effects on the filtering mass.

\begin{figure}
\includegraphics[width=0.9\columnwidth]{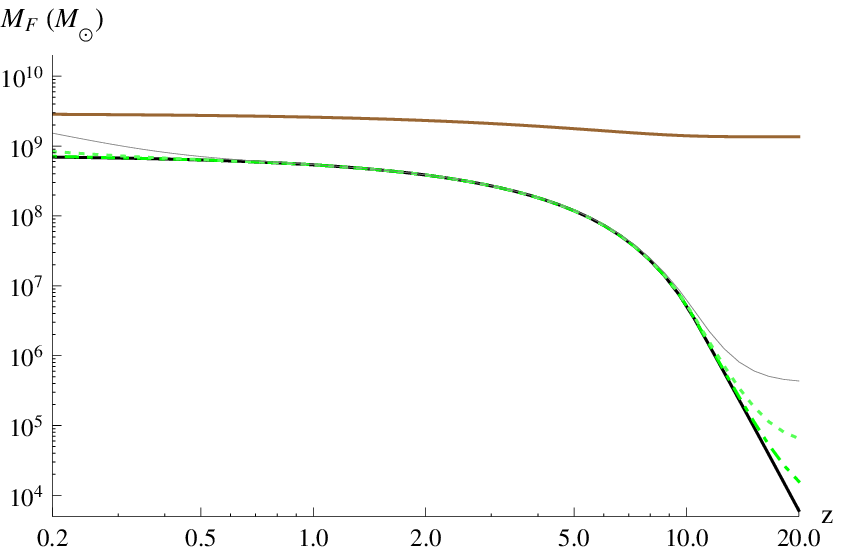}
\caption{
Variation of the filtering mass with redshift in the presence of a dipole-like ($p=3/2$) random magnetic field, taking into account amplification of the seed fields by IGM turbulence,  for $z_s=11$ and $z_r=8$. The bottom continuous (\textit{black}) curve corresponds to the $B_0=0$ case. The top continuous (\textit{brown}) curve corresponds to the CMB upper-limit $B_0\approx2.98\nG$ and $L_0=1\text{ Mpc}$. The other curves have $B_0=10\nG$ and, from bottom to top, 
$L_0=10^{2}\text{ pc}$ for the dash-dotted  (\textit{green}) curve;
$L_0=10^{2.5}\text{ pc}$ for the dotted (\textit{light-green}) curve;
$L_0=10^{3}\text{ pc}$ for the thin (\textit{gray}) curve.
}
\label{fig:filter8turb}
\end{figure}

\begin{figure}
\includegraphics[width=0.9\columnwidth]{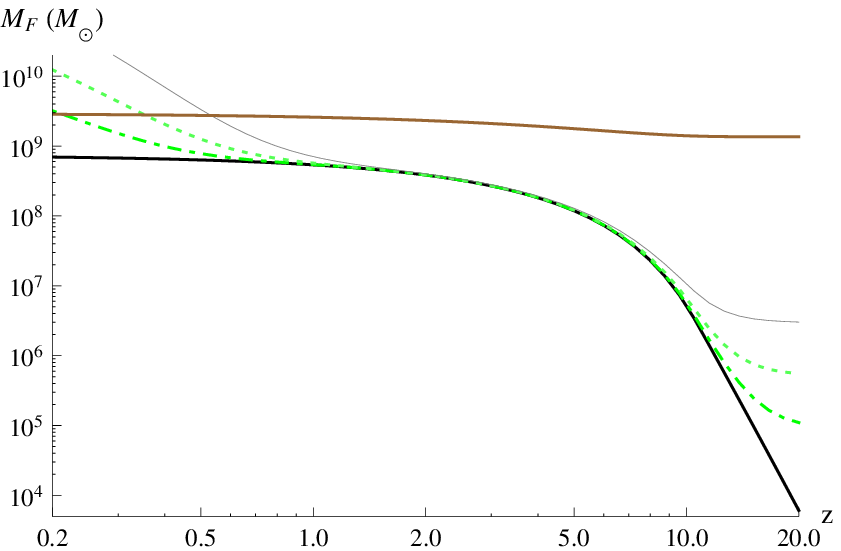}
\caption{
Variation of the filtering mass with redshift in the presence of a ring-like ($p=1$) random magnetic field, taking into account amplification of the seed fields by IGM turbulence, for $z_s=11$ and $z_r=8$. The bottom continuous (\textit{black}) curve corresponds to the $B_0=0$ case. The top continuous (\textit{brown}) curve corresponds to the CMB upper-limit $B_0\approx2.98\nG$ and $L_0=1\text{ Mpc}$. The other curves have $B_0=1\nG$ and, from bottom to top, 
$L_0=10^{2}\text{ pc}$ for the dash-dotted  (\textit{green}) curve;
$L_0=10^{3}\text{ pc}$ for the dotted (\textit{light-green}) curve;
$L_0=10^{4}\text{ pc}$ for the thin (\textit{gray}) curve.
}
\label{fig:filter8ringturb}
\end{figure}

Observations of the cosmic microwave background radiation (CMB) lead to an upper limit on the homogeneous  primordial magnetic field $B_{CMB}=2.98\nG$ (comoving) \citep{Yamazaki2010} with $L_0\sim 1\,\text{Mpc}$. This limit corresponds to the brown curve plotted in figures \ref{fig:filter8}, \ref{fig:filter8turb}, \ref{fig:filter8ring} and \ref{fig:filter8ringturb}.  There is, thus, a family of possible models to explain  the origin of cosmic magnetic fields in the early Universe that can create a difference in the filtering mass between  $10^4-10^{9.5} M_{\odot}$ and is in agreement with the CMB constraints.

The increase of the filtering mass due to the presence of magnetic fields is bigger before the reionization era, since the temperature, then,  contributes less to the total pressure. 

\begin{figure}
\includegraphics[width=0.9\columnwidth]{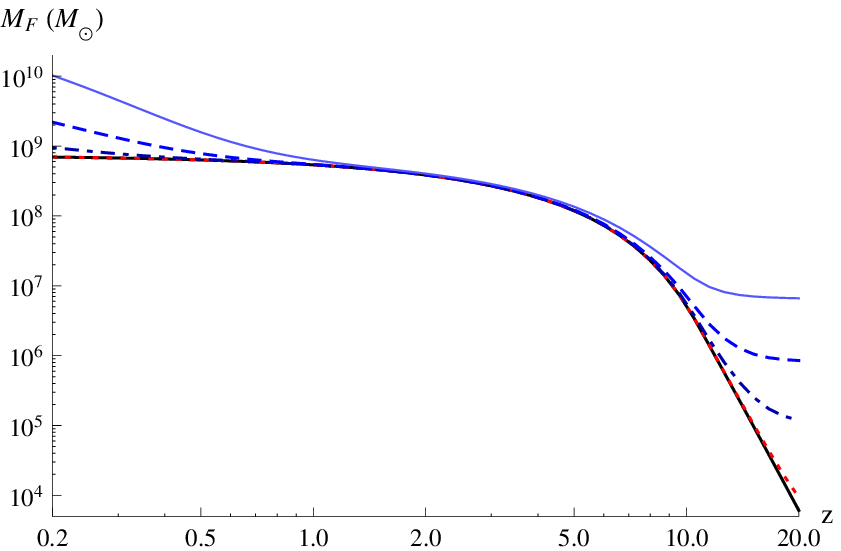}
\caption{Variation of the filtering mass with redshift in the presence of a dipole-like ($p=3/2$) random magnetic field,  taking into account amplification of the seed fields by IGM turbulence,   for $z_s=11$ and $z_r=8$. The continuous (\textit{black}) curve corresponds to the $B_0=0$ case. The other curves have $B_0=0.1\,\mu\text{G}$ and, from bottom to top, 
$L_0=10\text{ pc}$ for the dotted (\textit{red}) curve;
$L_0=10^2\text{ pc}$ for the dash-dotted (\textit{dark-blue}) curve;
$L_0=10^{2.5}\text{ pc}$ for the dashed (\textit{blue}) curve;
$L_0=10^{3}\text{ pc}$ for the thin (\textit{light-blue}) curve.}
\label{fig:filter7turb}
\end{figure}

\begin{figure}
\includegraphics[width=0.9\columnwidth]{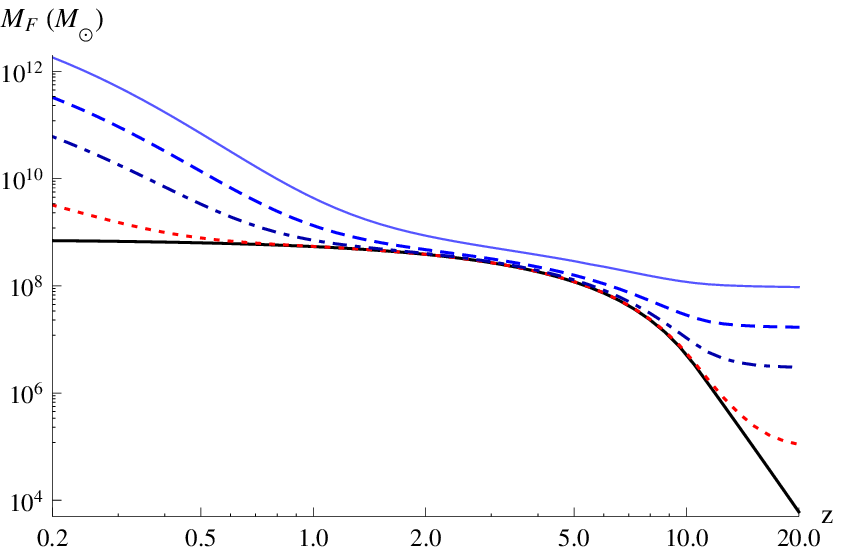}
\caption{
Variation of the filtering mass with redshift in the presence of a ring-like ($p=1$) random magnetic field, taking into account amplification of the seed fields by IGM turbulence,  for $z_s=11$ and $z_r=8$. The continuous (\textit{black}) curve corresponds to the $B_0=0$ case. The other curves have $B_0=0.1\,\mu\text{G}$ and, from bottom to top, 
$L_0=10\text{ pc}$ for the dotted (\textit{red}) curve;
$L_0=10^2\text{ pc}$ for the dash-dotted (\textit{dark-blue}) curve;
$L_0=10^{2.5}\text{ pc}$ for the dashed (\textit{blue}) curve;
$L_0=10^{3}\text{ pc}$ for the thin (\textit{light-blue}) curve.}
\label{fig:filter7ringturb}
\end{figure}

We also considered   that the seed field could have been amplified by effects of intergalactic turbulence  (as discussed in section \ref{sec:turb}). The evolution of the filtering mass considering this effect is shown in figures \ref{fig:filter7turb} and \ref{fig:filter8turb} for dipole-like fields and  \ref{fig:filter7ringturb} and  \ref{fig:filter8ringturb}  for ring-like fields. Comparing these figures with the previous ones, we note that the amplification  leads to an increase in the filtering mass only at small redshifts.

\section{Gas Fraction Content}\label{sec:gas}

From numerical simulations, \citet{Gnedin2000a} showed that the filtering mass determines the mass fraction of baryonic matter which can be found inside halos. Quantitatively, he found that the fraction, $f_{g}$, of the mass of the halo of total mass $M$, in the form of baryonic gas, can be approximated by the expression
\begin{equation}
 f_g \approx \frac{f_b}{\left[ 1+0.26 M_{F}(t)/M \right]^3}\label{eq:gashalo}
\end{equation}
where $f_{b}=\frac{\Omega_b}{\Omega_m}$ is the cosmic baryon to mass fraction.

\begin{figure}
\includegraphics[width=0.49\columnwidth]{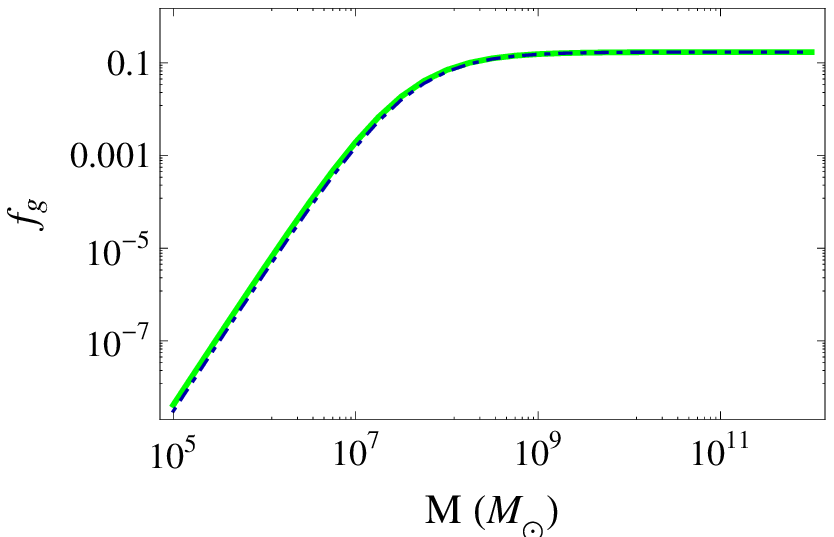}\hfill \includegraphics[width=0.49\columnwidth]{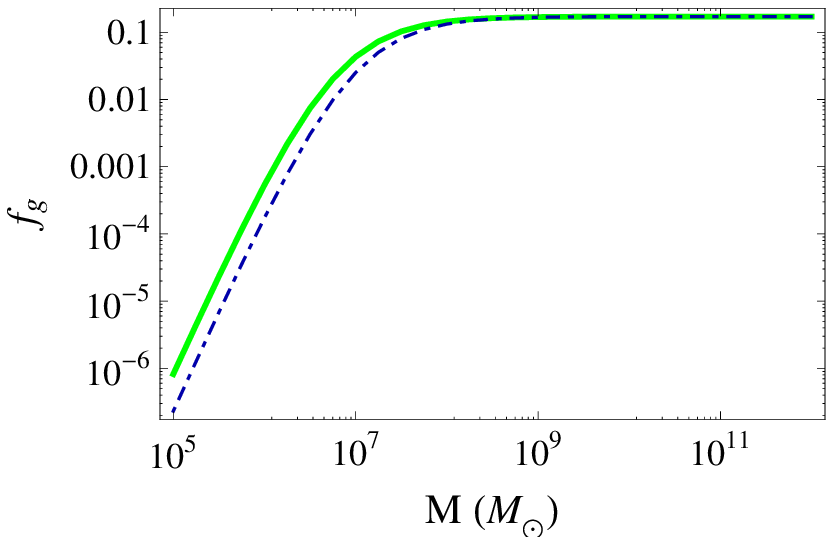}
\includegraphics[width=0.49\columnwidth]{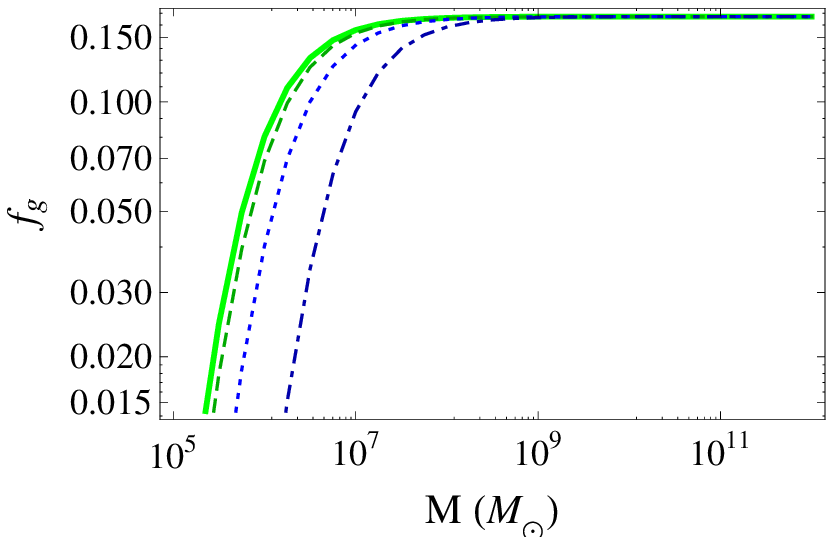}\hfill \includegraphics[width=0.49\columnwidth]{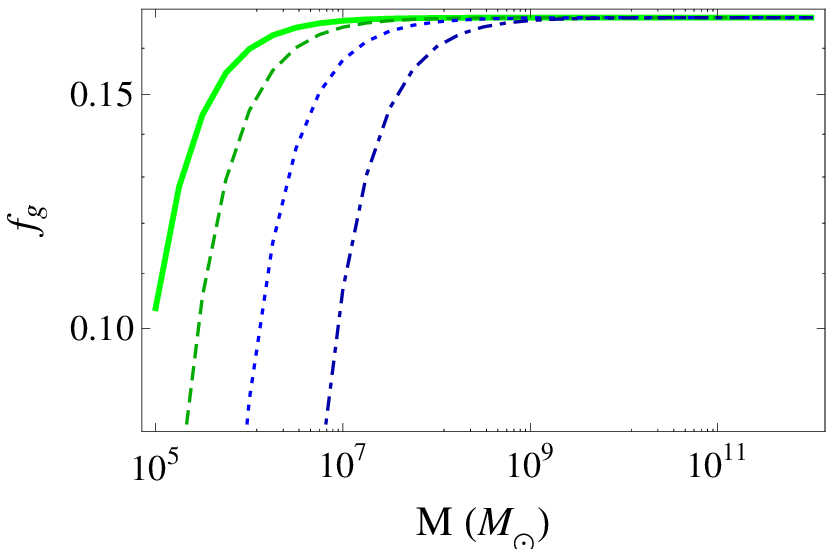}
\caption{Halo gas fraction, in the presence of a dipole-like field ($p=3/2$), as a function of halo mass at $z=3$, $z=6$, $z=9$ and $z=12$, in the \emph{top-left}, \emph{top-right}, \emph{bottom-left} and \emph{bottom-right} panels, respectively.  
The thick (\textit{green}) curve corresponds to $B_0=0$. For all other curves, $B_0=10^{-7}\text{ G}$. 
For the dashed (\textit{dark-green}) curve we have  $L_0=10\text{ pc}$,
for the dotted (\textit{blue}) curve,  $L_0=10^2\text{ pc}$, 
for the dash-dotted (\textit{dark-blue}) curve,  $L_0=10^3\text{ pc}$.
}\label{fig:fgas}
\end{figure}

\begin{figure}\includegraphics[width=0.49\columnwidth]{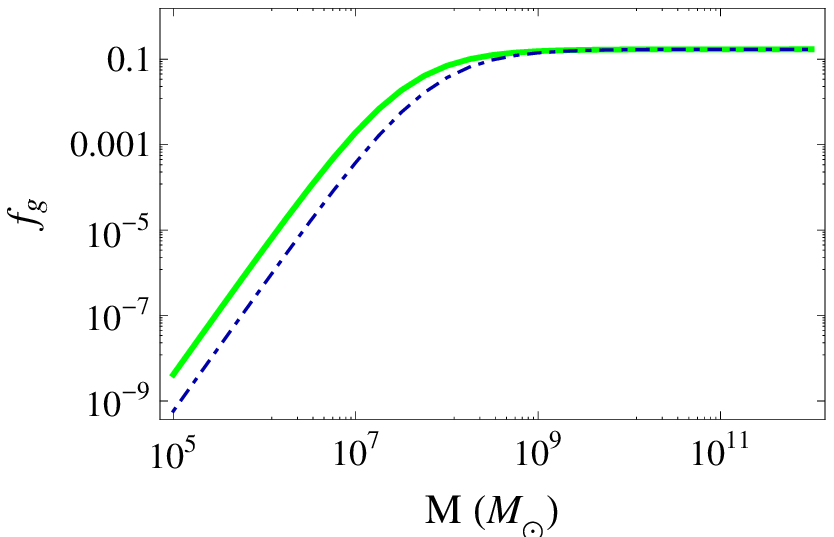}\hfill \includegraphics[width=0.49\columnwidth]{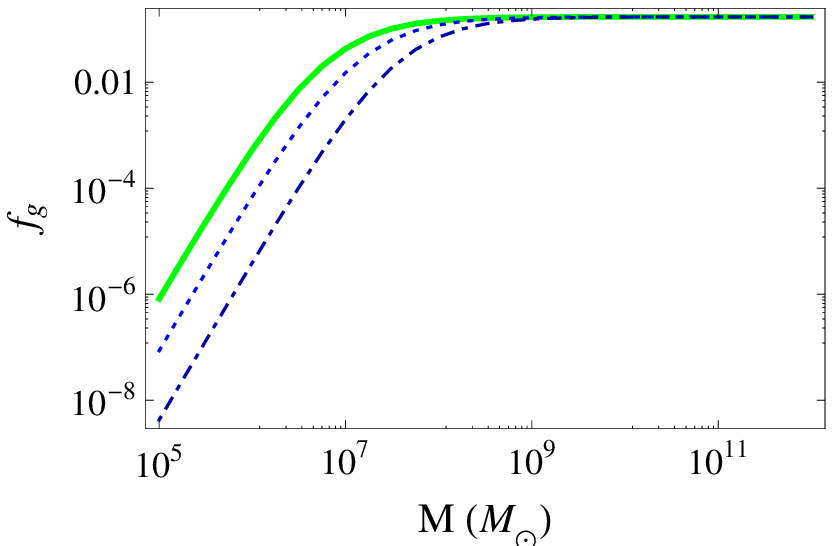}
\includegraphics[width=0.49\columnwidth]{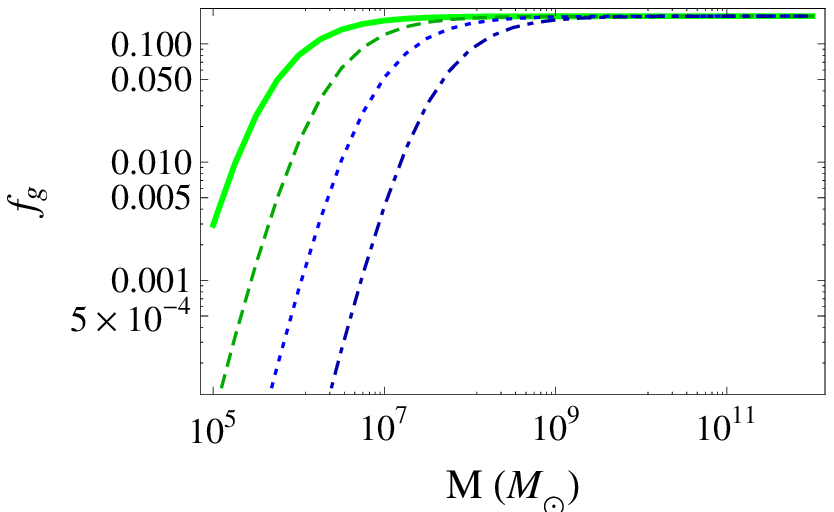}\hfill \includegraphics[width=0.49\columnwidth]{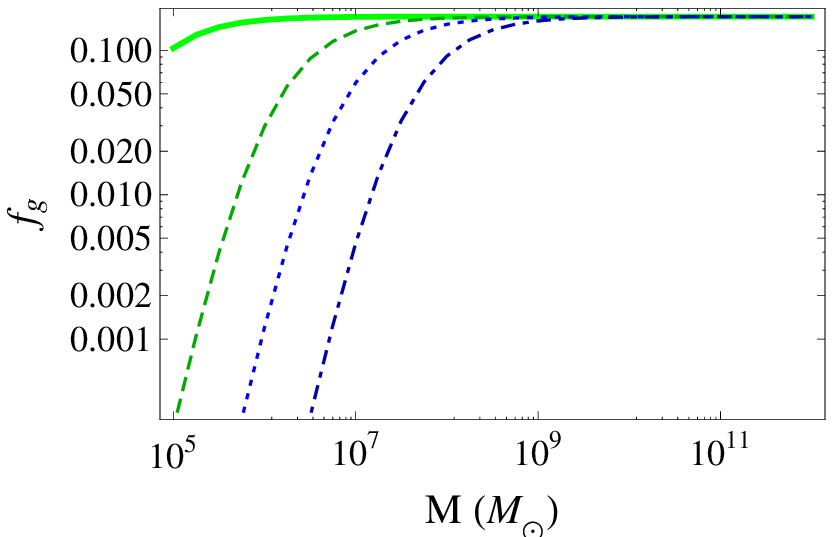}
\caption{Halo gas fraction, in the presence of a ring-like field ($p=1$), as a function of halo mass at $z=3$, $z=6$, $z=9$ and $z=12$, in the \emph{top-left}, \emph{top-right}, \emph{bottom-left} and \emph{bottom-right} panels, respectively.  
The thick (\textit{green}) curve corresponds to $B_0=0$. For all other curves, $B_0=10^{-7}\text{ G}$. 
For the dashed (\textit{dark-green}) curve we have  $L_0=10\text{ pc}$,
for the dotted (\textit{blue}) curve,  $L_0=10^2\text{ pc}$, 
for the dash-dotted (\textit{dark-blue}) curve,  $L_0=10^3\text{ pc}$.
}\label{fig:fgasring}
\end{figure}

Using our expression for the magnetic Jeans mass, we evaluate the gas fraction for different values of $B_0$ and $L_0$. We also considered two possible geometries for the seed field and the possibility of the seed field to be amplified by IGM turbulence.  The results are presented in figures \ref{fig:fgas}, 
 \ref{fig:fgasring}.

As expected, we find a dramatic decrease in the gas fraction for small mass halos, due to the presence of the magnetic field. The fraction of gas can be changed by 2-3 orders of magnitude  at high redshift depending on the value of $B_0$ and the coherence length of the primordial magnetic field, $L_0$.

\section{Conclusions}\label{sec:conc}

We modified the Jeans mass in order to take into account  the presence of  random primordial magnetic fields (RPMF) in the form of randomly oriented cells with dipole and ring-like structures. From this modified Jeans mass, we obtained the filtering mass and the baryonic gas fraction of a dark matter halo. We showed that, depending on the magnetogenesis  model, which determines $B_0$ and $L_0$,  both the Jeans mass and the baryonic gas fraction can change by orders of magnitude. 
We found, for example, for a comoving $B_0=0.1\muG$, and a reionization epoch that starts at $z_s=11$ and ends at $z_e=8$, for $L_0=100\,\text{pc}$ at $z=12$, the $f_g$ becomes severely depressed for $M<10^7\msun$, whereas for $B_0=0$ the $f_g$ becomes severely depressed only for much smaller masses, $M<10^5\msun$. 

Since it is very difficult to make observations of intergalactic magnetic fields at high redshifts, and the constraints imposed by CMB measurements are not very restrictive, we suggest the possibility to add new constraints on a family of models for the primordial magnetic field, by following the  redshift evolution of the filtering mass of galaxies.  

We also calculated the modified baryonic gas fraction that can also be used as an indirect observable to help us to understand the origin and structure of cosmic magnetic fields.

\section*{Acknowledgments}

R.S.S. thanks the Brazilian agency FAPESP for financial support (2009/06770-2). 
L.F.S.R. thanks the Brazilian agency CNPq for financial support (142394/2006-8). 
R.O. thanks the Brazilian agencies FAPESP (06/56213-9) and  CNPq (300414/82-0) for partial
support.  We also thank all comments of the anonymous referee that helped us to improve the present work.

\bsp

\label{lastpage}

\begin{thebibliography}{}

\bibitem[\protect\citeauthoryear{{Ahonen} \& {Enqvist}}{{Ahonen} \&
  {Enqvist}}{1998}]{Ahonen1997}
{Ahonen} J.,  {Enqvist} K.,  1998, Phys. Rev. D, 57, 664

\bibitem[\protect\citeauthoryear{Barkana \& Loeb}{Barkana \& Loeb}{2001}]{rev}
Barkana R.,  Loeb A.,  2001, Physics Reports, 349, 125

\bibitem[\protect\citeauthoryear{Barkana \& Loeb}{Barkana \& Loeb}{2002}]{bl02}
Barkana R.,  Loeb A.,  2002, ApJ, 578, 1

\bibitem[\protect\citeauthoryear{{Baym}, {B{\"o}deker} \& {McLerran}}{{Baym}
  et~al.}{1996}]{Baym1996}
{Baym} G.,  {B{\"o}deker} D.,    {McLerran} L.,  1996, Phys. Rev. D, 53, 662

\bibitem[\protect\citeauthoryear{Biermann}{Biermann}{1950}]{Biermann1950}
Biermann L.,  1950, Zeitschrift Naturforschung Teil A, 5

\bibitem[\protect\citeauthoryear{Cen}{Cen}{2003}]{Cen}
Cen R.,  2003, ApJ, 591, L5

\bibitem[\protect\citeauthoryear{{Cheng} \& {Olinto}}{{Cheng} \&
  {Olinto}}{1994}]{ChengOlinto}
{Cheng} B.,  {Olinto} A.~V.,  1994, Phys. Rev. D, 50, 2421

\bibitem[\protect\citeauthoryear{Clarke, Kronberg \& Boehringer}{Clarke
  et~al.}{2001}]{Clarke2001}
Clarke T.~E.,  Kronberg P.~P.,    Boehringer H.,  2001, ApJ, 547, L111

\bibitem[\protect\citeauthoryear{Davies \& Widrow}{Davies \&
  Widrow}{2000}]{Davies00}
Davies G.,  Widrow L.~M.,  2000, The Astrophysical Journal, 540, 755

\bibitem[\protect\citeauthoryear{de Souza \& Opher}{de~Souza \&
  Opher}{2008}]{Souza2008}
de Souza R.,  Opher R.,  2008, Physical Review D, 77, 043528

\bibitem[\protect\citeauthoryear{{de Souza} \& {Opher}}{{de Souza} \&
  {Opher}}{2010a}]{souza10}
{de Souza} R.~S.,  {Opher} R.,  2010a, Journal of Cosmology and Astro-Particle
  Physics, 2, 22

\bibitem[\protect\citeauthoryear{{de Souza} \& {Opher}}{{de Souza} \&
  {Opher}}{2010b}]{Souza2010}
{de Souza} R.~S.,  {Opher} R.,  2010b, Physical Review D, 81, 067301

\bibitem[\protect\citeauthoryear{{Enqvist} \& {Olesen}}{{Enqvist} \&
  {Olesen}}{1993}]{Enqvist1993}
{Enqvist} K.,  {Olesen} P.,  1993, Physics Letters B, 319, 178

\bibitem[\protect\citeauthoryear{Furlanetto \& Oh}{Furlanetto \&
  Oh}{2006}]{Furlanetto06}
Furlanetto S.~R.,  Oh S.~P.,  2006, ApJ, 652, 849

\bibitem[\protect\citeauthoryear{Gnedin}{Gnedin}{2000}]{Gnedin2000a}
Gnedin N.~Y.,  2000, ApJ, 542, 535

\bibitem[\protect\citeauthoryear{Gnedin \& Hui}{Gnedin \&
  Hui}{1998}]{Gnedin1997}
Gnedin N.~Y.,  Hui L.,  1998, MNRAS, 296, 44

\bibitem[\protect\citeauthoryear{Grasso \& Rubinstein}{Grasso \&
  Rubinstein}{2001}]{Grasso2001}
Grasso D.,  Rubinstein H.,  2001, Physics Reports, 348, 163

\bibitem[\protect\citeauthoryear{Haiman \& Holder}{Haiman \&
  Holder}{2003}]{HH03}
Haiman Z.,  Holder G.~P.,  2003, ApJ, 595, 1

\bibitem[\protect\citeauthoryear{Han}{Han}{2008}]{han08}
Han J.~L.,  2008, Nuclear Physics B - Proceedings Supplements, 175, 62

\bibitem[\protect\citeauthoryear{Hanayama, Takahashi, Kotake, Oguri, Ichiki \&
  Ohno}{Hanayama et~al.}{2005}]{Hanayama2005}
Hanayama H.,  Takahashi K.,  Kotake K.,  Oguri M.,  Ichiki K.,    Ohno H.,
  2005, ApJ, 633, 941

\bibitem[\protect\citeauthoryear{{Hindmarsh} \& {Everett}}{{Hindmarsh} \&
  {Everett}}{1998}]{Hindmarsh1998}
{Hindmarsh} M.,  {Everett} A.,  1998, Physical Review D, 58, 103505

\bibitem[\protect\citeauthoryear{{Hogan}}{{Hogan}}{1983}]{Hogan1983}
{Hogan} C.~J.,  1983, Physical Review Letters, 51, 1488

\bibitem[\protect\citeauthoryear{Ichiki, Takahashi, Ohno, Hanayama \&
  Sugiyama}{Ichiki et~al.}{2006}]{Ichiki2006}
Ichiki K.,  Takahashi K.,  Ohno H.,  Hanayama H.,    Sugiyama N.,  2006, Sci.,
  311, 827

\bibitem[\protect\citeauthoryear{Iliev, Scannapieco, Martel \& Shapiro}{Iliev
  et~al.}{2003}]{iliev2}
Iliev I.~T.,  Scannapieco E.,  Martel H.,    Shapiro P.~R.,  2003, MNRAS, 341,
  81

\bibitem[\protect\citeauthoryear{Iliev, Scannapieco \& Shapiro}{Iliev
  et~al.}{2005}]{iss05}
Iliev I.~T.,  Scannapieco E.,    Shapiro P.~R.,  2005, ApJ, 624, 491

\bibitem[\protect\citeauthoryear{Kravtsov, Gnedin \& Klypin}{Kravtsov
  et~al.}{2004}]{Kravtsov2004}
Kravtsov A.~V.,  Gnedin O.~Y.,    Klypin A.~A.,  2004, ApJ, 609, 482

\bibitem[\protect\citeauthoryear{Kronberg}{Kronberg}{1994}]{Kronberg1994}
Kronberg P.~P.,  1994, Reports on Progress in Physics, 57, 325

\bibitem[\protect\citeauthoryear{Kuhlen, Madau \& Montgomery}{Kuhlen
  et~al.}{2006}]{Kuhlen}
Kuhlen M.,  Madau P.,    Montgomery R.,  2006, ApJ, 637, L1

\bibitem[\protect\citeauthoryear{Kulsrud \& Zweibel}{Kulsrud \&
  Zweibel}{2008}]{Kulsrud2008}
Kulsrud R.~M.,  Zweibel E.~G.,  2008, Reports on Progress in Physics, 71,
  046901

\bibitem[\protect\citeauthoryear{{Lagan{\'a}}, {de Souza} \&
  {Keller}}{{Lagan{\'a}} et~al.}{2010}]{lagana}
{Lagan{\'a}} T.~F.,  {de Souza} R.~S.,    {Keller} G.~R.,  2010, A\&A, 510, A76

\bibitem[\protect\citeauthoryear{{Macci{\`o}}, {Kang}, {Fontanot},
  {Somerville}, {Koposov} \& {Monaco}}{{Macci{\`o}} et~al.}{2010}]{Maccio2009}
{Macci{\`o}} A.~V.,  {Kang} X.,  {Fontanot} F.,  {Somerville} R.~S.,  {Koposov}
  S.,    {Monaco} P.,  2010, MNRAS, 402, 1995

\bibitem[\protect\citeauthoryear{McQuinn, Lidz, Zahn, Dutta, Hernquist \&
  Zaldarriaga}{McQuinn et~al.}{2007}]{mcquinn07}
McQuinn M.,  Lidz A.,  Zahn O.,  Dutta S.,  Hernquist L.,    Zaldarriaga M.,
  2007, MNRAS, 377, 1043

\bibitem[\protect\citeauthoryear{Maeda, Kitagawa, Kobayashi \& Shiromizu}{Maeda
  et~al.}{2009}]{Maeda2009}
Maeda S.,  Kitagawa S.,  Kobayashi T.,    Shiromizu T.,  2009, Classical and
  Quantum Gravity, 26, 135014

\bibitem[\protect\citeauthoryear{Miranda, Opher \& Opher}{Miranda
  et~al.}{1998}]{Miranda1998}
Miranda O.,  Opher M.,    Opher R.,  1998, MNRAS, 301, 547

\bibitem[\protect\citeauthoryear{Naoz \& Barkana}{Naoz \& Barkana}{2008}]{NB08}
Naoz S.,  Barkana R.,  2008, MNRAS, 385, L63

\bibitem[\protect\citeauthoryear{Naoz, Noter \& Barkana}{Naoz
  et~al.}{2006}]{NNB}
Naoz S.,  Noter S.,    Barkana R.,  2006, MNRAS, 373, L98

\bibitem[\protect\citeauthoryear{Padmanabhan}{Padmanabhan}{2002}]{Padmanabhan3}
Padmanabhan T.,  2002, {Theoretical Astrophysics: Volume III: Galaxies and
  Cosmology}.
Cambridge University Press

\bibitem[\protect\citeauthoryear{Peebles}{Peebles}{1967}]{Peebles1967}
Peebles P. J.~E.,  1967, ApJ, 147, 859

\bibitem[\protect\citeauthoryear{Rees \& Reinhardt}{Rees \&
  Reinhardt}{1972}]{Rees1972}
Rees M.~J.,  Reinhardt M.,  1972, A\&A, 19

\bibitem[\protect\citeauthoryear{{Rodrigues}, {de Souza} \&
  {Opher}}{{Rodrigues} et~al.}{2010}]{Rodrigues2010}
{Rodrigues} L.~F.~S.,  {de Souza} R.~S.,    {Opher} R.,  2010, MNRAS, 406, 482

\bibitem[\protect\citeauthoryear{{Ryu}, {Kang}, {Cho} \& {Das}}{{Ryu}
  et~al.}{2008}]{Ryu2008}
{Ryu} D.,  {Kang} H.,  {Cho} J.,    {Das} S.,  2008, Science, 320, 909

\bibitem[\protect\citeauthoryear{Schleicher, Banerjee \& Klessen}{Schleicher
  et~al.}{2008}]{Schleicher2008}
Schleicher D. R.~G.,  Banerjee R.,    Klessen R.~S.,  2008, Phys. Rev. D, 78,
  083005

\bibitem[\protect\citeauthoryear{Shapiro, Ahn, Alvarez, Iliev, Martel \&
  Ryu}{Shapiro et~al.}{2006}]{Shapiro06}
Shapiro P.~R.,  Ahn K.,  Alvarez M.~A.,  Iliev I.~T.,  Martel H.,    Ryu D.,
  2006, ApJ, 646, 681

\bibitem[\protect\citeauthoryear{Takahashi, Ichiki, Ohno \& Hanayama}{Takahashi
  et~al.}{2005}]{Takahashi2005}
Takahashi K.,  Ichiki K.,  Ohno H.,    Hanayama H.,  2005, Physical Review
  Letters, 95, 4

\bibitem[\protect\citeauthoryear{Takahashi, Ichiki, Ohno, Hanayama \&
  Sugiyama}{Takahashi et~al.}{2006}]{Takahashi2006}
Takahashi K.,  Ichiki K.,  Ohno H.,  Hanayama H.,    Sugiyama N.,  2006,
  Astronomische Nachrichten, 327, 410

\bibitem[\protect\citeauthoryear{Tashiro \& Sugiyama}{Tashiro \&
  Sugiyama}{2005}]{Tashiro2005}
Tashiro H.,  Sugiyama N.,  2005, MNRAS, 368, 965

\bibitem[\protect\citeauthoryear{Wasserman}{Wasserman}{1978}]{Wasserman1978}
Wasserman I.,  1978, ApJ, 224, 337

\bibitem[\protect\citeauthoryear{Widrow}{Widrow}{2002}]{Widrow2002}
Widrow L.~M.,  2002, Reviews of Modern Physics, 74, 775

\bibitem[\protect\citeauthoryear{Wyithe \& Loeb}{Wyithe \& Loeb}{2003}]{WL03}
Wyithe J. S.~B.,  Loeb A.,  2003, ApJ, 588, L69

\bibitem[\protect\citeauthoryear{Yamazaki, Ichiki, Kajino \& Mathews}{Yamazaki
  et~al.}{2010}]{Yamazaki2010}
Yamazaki D.~G.,  Ichiki K.,  Kajino T.,    Mathews G.~J.,  2010, Phys. Rev. D,
  81, 023008

\bibitem[\protect\citeauthoryear{Zweibel \& Heiles}{Zweibel \&
  Heiles}{1997}]{Zweibel1997}
Zweibel E.~G.,  Heiles C.,  1997, Nat., 385, 131

\end{thebibliography}
\end{document}